\documentclass[sort&compress,10pt]{iopart}
\pdfoutput=1
\usepackage{iopams}

\usepackage[T1]{fontenc} %
\usepackage[english]{babel}
\usepackage{bm}
\usepackage[usenames,dvipsnames]{color}
\usepackage{cite}
\usepackage{graphicx}
\usepackage[colorlinks=true,linkcolor=blue,citecolor=blue,urlcolor=blue,pdfborder={0 0 0}]{hyperref}
\usepackage[utf8]{inputenx}
\usepackage{xspace}
\usepackage{siunitx}
\usepackage[scaled=0.9]{helvet}      %
\usepackage{mathptmx}    %

\newcommand\potfit{\textit{potfit}\xspace}
\newcommand{\AI}{\textit{ab initio}\xspace}
\newcommand{\thz}{\tera\hertz}

\begin{document}

\title[\sffamily Classical potentials from \AI data: a review of \potfit]%
{\sffamily Classical interaction potentials for diverse materials from \AI data: a review of \potfit}

\author{\sffamily Peter Brommer$^1$ and Alexander Kiselev$^2$ and Daniel Schopf$^3$ and Philipp Beck$^4$ and Johannes Roth$^2$ and
Hans-Rainer Trebin$^2$}

\address{$^1$Warwick Centre for Predictive Modelling, School of Engineering, and Centre for Scientific Computing, University of Warwick, Library Road, Coventry CV4 7AL, UK}
\address{$^2$Institut für Funktionelle Materie und Quantentechnologien (FMQ), Universität Stuttgart, Pfaffenwaldring 57, 70569 Stuttgart, Germany}
\address{$^3$Cinemo GmbH, Kaiserstraße 72, 76133 Karlsruhe, Germany}
\address{$^4$Semcon Bad Friedrichshall GmbH, Bergrat-Bilfinger-Straße 5, 74177 Bad Friedrichshall, Germany}
\ead{\rmfamily \href{mailto:p.brommer@warwick.ac.uk}{\rmfamily p.brommer@warwick.ac.uk}}

\begin{abstract}
  Force matching is an established technique to generate effective potentials for molecular dynamics simulations from first-principles data.
  This method has been implemented in the open source code \potfit.
  Here, we present a review of the method and describe the main features of the code.
  Particular emphasis is placed on the features added since the initial release: interactions represented by analytical functions, differential evolution as optimization method, and a greatly extended set of interaction models.
  Beyond the initially present pair and embedded-atom method potentials, \potfit can now also optimize angular dependent potentials, charge and dipolar interactions, and electron-temperature-dependent potentials.
  We demonstrate the functionality of these interaction models using three example systems: phonons in type I clathrates, fracture of $\alpha$-alumina, and laser-irradiated silicon.
\end{abstract}

\submitto{\MSMSE}

\section{Introduction}
\label{sec:intro}
The enormous increase of available computational resources in the past decades has powered the success of molecular dynamics (MD), i.e., the computer simulation of the trajectories of atoms (and sometimes molecules) by numerically integrating Newton's equations of motion, in materials simulation.
Molecular dynamics studies can routinely comprise millions of time steps for millions of atoms, and up to several trillions in a recent proof-of-concept \cite{Eckhardt:2013:1}.
Such large numbers of particles and integration steps require an efficient way to evaluate the interactions of a system.
Even with the vast advances in first-principles methods like density functional theory (DFT), this can only be provided with classical effective interaction potentials or force fields.
These represent the energy (and as its gradients the internal forces) of a system as a function of only the positions of the atoms, eliminating all electronic degrees of freedom.
The calculation is further simplified by the fact that in general only two-, three- or rarely four-body terms are used to specify the system energy.
These calculations are also efficiently parallelized, which is a prerequisite for such large systems.
Typical applications that require millions of atoms are simulations of fracture \cite{Roesch:2009:56002}, dislocations \cite{Bitzek:2013:1394} or  laser ablation \cite{Sonntag:2010:77}.

In contrast to \AI methods, which are, at least in principle, literally parameter-free, interaction models contain parameters, which the simulator must choose appropriately.
Their values are used to calculate the energies and forces, which determine the trajectories of the particles and thus all properties resulting from a MD simulation.
Also, these parameters are in general specific not only to an element, but to the complete set of atomic species involved in a simulation---or even a single chemical composition \cite{Becker:2013:277}.
A researcher wishing to undertake MD simulations in a certain system can peruse the literature in search of the appropriate parameter set.
There are also a number of repositories that host potentials, e.g.\ the OpenKIM project\footnote{\url{https://openkim.org}} \cite{Tadmor:2011:17,Tadmor:2013:298} or NIST's Interatomic Potentials Repository Project\footnote{\url{http://www.ctcms.nist.gov/potentials}} \cite{Becker:2013:277}.
At the time of writing, those two repositories contain interaction potentials for many of the elements, but only for around 50 binary and 13 (!) ternary systems.
In contrast, the AFLOWLIB project \cite{Curtarolo:2012:227}, a first-principles high-throughput database of materials, currently contains information on over 500 binary compounds alone.
Thus, one of the major limitations of molecular dynamics is the availability of suitable interaction potentials.
Consequently, there is a sustained demand to determine effective potentials for more and more alloy systems -- and for tools to assist researchers in doing this.

The ``traditional'' approach to this problem was to fit the potential to experimental data or \AI calculated parameters such as lattice constants, bulk moduli and elastic constants.
However, for complex systems such data might be scarce, unreliable or even unavailable, which could drive the users to resort to model potentials with little physical justification.
Here, the force matching method  \cite{Ercolessi:1994:583} offers a way on: It uses a large database of \AI calculated reference data, such as forces and energies, to optimize a potential, defined as a set of independent parameters.
This is based on the idea, that if a potential yields correct energies and forces acting on the individual atoms, it can also produce the correct collective dynamics.
The dynamics in turn are responsible for measured properties such as phase transitions, free energy differences or elastic properties.

The open source program \potfit\footnote{\url{http://potfit.sourceforge.net/}} \cite{Brommer:2006:753,Brommer:2007:295} has been developed as a highly flexible implementation of the force matching method, in both the potential forms used and the optimization schemes to arrive at the final parameters.
Since the original publication, the functionality of the code has been considerably expanded beyond the pair and embedded atom method (EAM) potentials presented there.
While the initial set of interaction models cover many ``simple'' metals  (i.e., without any directional bonding), angular-dependent potentials (ADP) \cite{Mishin:2005:4029} can reproduce certain covalent aspects in a metallic bonding environment.
Polarizable potentials of the Tangney-Scandolo (TS) type \cite{Tangney:2002:8898} extend \potfit to force-match oxides.
Electron temperature-dependent potentials \cite{Bennemann:2004:995,Recoules:2006:55503} are significant wherever the electronic temperature of the system varies strongly; this is most significant for simulation of laser--matter interactions.

In this article, we provide an overview of the force matching method and then review the \potfit code in section \ref{sec:algo}, with particular emphasis on the features introduced since the original publications:
Differential evolution as optimization method (section \ref{sec:de}), analytic potential representation (section \ref{sec:potent-repr}), and the new potential models mentioned above (section \ref{sec:pot-adp}).
In section \ref{sec:test}, we present three applications that make use of the new components: Phonons in clathrates (section \ref{sec:adp}), fracture of $\alpha$-alumina (section \ref{sec:dipole}) and laser irradiation of silicon (section \ref{sec:temp-depend-potent}).

\section{Methods and models}
\label{sec:algo}

\subsection{Force matching}
\label{sec:force-matching}

\subsubsection{Interaction potentials from first principles}
\label{sec:inter-potent-from}

The force matching method was first described around 20 years ago by Ercolessi and Adams \cite{Ercolessi:1994:583}.
It makes use of the ever increasing availability and predictive power of first-principles simulations, which---while by far not fast enough for a straight \AI simulation---can be used to parameterize an interaction potential.
To this end, it directly uses microscopic quantities from first principles simulations, such as internal forces acting on an individual atom, as reference data in the parameterization process.
This allows the determination of effective potentials even in cases, where there is insufficient experimental data to fit a sufficiently complex interaction model.
The main motivation behind the force matching method is that a potential that yields forces correctly will reproduce correct dynamics, which then in turn leads to correct macroscopic quantities determined from MD simulations.

Fitting a potential is an optimization process.
Force matching uses a weighted sum of squares $Z$, defined as the deviations from the \AI reference data, as the target function for the optimization.
A set of parameters, called $\balpha$, is adjusted to reproduce the reference data as accurate as possible.
The target function $Z(\balpha)$ is defined as
\begin{eqnarray}
  &Z(\balpha) &= Z_D(\balpha) + Z_C(\balpha), \label{eqn:sumofsquares} \\
  \mathrm{with} \quad &Z_D(\balpha) &= \sum_{i=0}^mu_i(S_i(\balpha)-S_i^0)^2 \\
  \mathrm{and} \quad &Z_C(\balpha) &= \sum_{r=0}^{N_C}w_r(A_r(\balpha)-A_r^0)^2.
\end{eqnarray}
The deviations $Z_D$ are made up from $m$ individual contributions $S$, such as forces, energies and stresses, multiplied by weighting factors $u_i$.
An additional term, $Z_C$, can be used to put constraints with weights $w_r$ on the optimization process, e.g.\ fixing gauge degrees of freedom.
The number $N_C$ of constraints treated in this way is typically dependent only on the potential type and not on the amount of reference information.
The quantities with a superscript $0$ are \AI calculated values.
It should be noted that the fitting process does not enforce any physical meaning of any parameter in the set $\balpha$ -- not even when the original description of a potential model provides one (like for example calling the proportionality constants in a Coulomb $r^{-1}$ interaction ``charges'').
They are rather chosen empirically to best reproduce the reference data, which are the only quantities with a direct physical meaning.

Since the inaugural publication, force matching has been used to parameterize countless interaction potentials.
The method has also been implemented in a number of codes that are now available under an open-source licence.
\emph{Potfit} \cite{Brommer:2007:295} was first presented in 2005 at the 9th International Conference for Quasicrystals \cite{Brommer:2006:753} as a method to determine potentials for quasicrystal-related structures, where experimental data is too scarce and unreliable to fit potentials.
As the main focus of this review, it will be discussed in more detail in section \ref{sec:force-matching-potfit} below.
ForceFit\cite{Waldher:2010:2307} was presented in 2010.
It provides a graphical user interface (GUI) to the force matching method and can fit analytic potentials with a particular emphasis on covalent interactions.
In 2013, ForceBalance \cite{Wang:2013:452} was introduced to parameterize force fields with a wide variety of functional forms.
It integrates the force matching method with an iterative improvement of reference structure selection and uses a local minimizer to optimize the parameters.
The selection of force field types (i.e., functional forms) integrated in ForceFit and ForceBalance demonstrate that these codes were designed for applications in organic chemistry, particularly aqueous solutions, whereas \potfit caters primarily to solid-state systems.
For a current review of the merits and shortcomings of the force matching method, see \cite{Masia:2014:1036}.

The standard force matching method is a sequential multiscale method, where first-principles simulations and classical MD simulations are performed separately, with information from one used to parameterize the other.
In contrast, hybrid quantum/classical (QM/MM) simulations (see e.g.~\cite{Bernstein:2009:26501} for a review) run both simulations at the same time, either by coupling the methods spatially (e.g.~\cite{Choly:2005:94101}), or by obtaining classical potential correction on the fly (e.g.~\cite{Csanyi:2004:175503,Li:2015:96405}).
The clear advantage of sequential simulations is that it allows to use dedicated tools for each multiscale layer.
These \AI and MD programs are well-established and highly optimized to their respective task.

There is no standard routine to select reference structures for force matching, however, there are empirical approaches that have proven successful in the past.
The general guideline is that the training data set should represent the local environments that will appear in MD simulations with the new potential reasonably well.
One way to achieve that is to use snapshots of a MD simulation, either \AI-MD or using an intermediate potential (or a combination of both \cite{Wang:2010:231101}).
This procedure can be iterated, i.e., the classical MD snapshots are updated with configurations from a trajectory determined with the next generation potential.
The reference data (forces, stresses and energies) is then calculated in these snapshots using first principles methods.
This data set can be complemented by configurations to cover specific situations (e.g.\ strained structures).
For some special-purpose potentials, the trajectory snapshots can be forgone entirely in favour of specifically targeted reference structures.
For example, a force-matched potential used to optimize complex ground state configurations would primarily contain low-temperature structures representing relevant structural motives \cite{Mihalkovic:2006:519}.

The selection of reference configurations also affects a central issue of force-matched potentials: transferability.
Can a potential be used under conditions that were not represented in the reference data?
For example, potentials for alloys that are based on DFT calculations of one particular composition (and do not use the pure constituents in the database) can not be expected to be used far from this composition.
This implies that force matching allows the creation of interactions that are tailored to a very particular situation.
On the other hand, a wider training set will allow for more flexible potentials.
In any case, a potential should be validated against information not directly used in the training data, but relevant for the desired application.
This can either be forces, stresses and energies from a test set, or quantities derived from more involved calculations such as elastic constants.

Following a rigorous testing regime will also help identify cases of overfitting, which may occur when there are too many adjustable parameters.
Then transferability may be lost completely, as the optimal solution may reproduce only the reference data, performing badly on configurations that are only slightly different.
Monitoring the success of the potential on a test data set will spot this behaviour and allow countermeasures.
It should however be kept in mind that even the best classical potential will most certainly have some limitations; the move from first-principles to an effective interaction incurs a loss of generality.
For this reason, creators and users of effective potentials must be aware of the capabilities and deficiencies of the interaction model.
As a consequence, all published force-matched potentials should be accompanied by detailed information on the selection of reference configurations used for generating it.

The force matching method is subtly different from other ways to derive atomistic force fields in relying strongly on large quantities of easily calculated data (forces, stresses and energies cost as much as a single MD step).
Alternative approaches aim to reproduce certain physical or chemical properties, such as for example formation heats, bond lengths and bond angles in the ReaxFF bond-order potential \cite{Duin:2001:9396}.
\footnote{Here, ReaxFF refers to the particular parameter set of this bond-order model.
  In principle, force matching could be used to determine a distinct (force-matched) parameter set. }
While this potential strives to be fully transferable (i.e., work for a certain atomic species under any circumstance), this strategy reaches its limits when the quantities required for the optimization process are no longer readily available.
For example, this may be the case for complex metallic alloys, where the atomic positions are not exactly known from experiment.
There, \AI-MD combined with an iterative refinement of classical MD trajectories may be used to provide the reference information for a force matching run.
This could then be used to determine a force-matched potential with the usual limitations on transferability -- which however typically uses ``simpler'' potentials with lower computational cost.

\subsubsection{Force matching with \potfit}
\label{sec:force-matching-potfit}

The \potfit package  \cite{Brommer:2007:295,Brommer:2006:753} is a flexible open source implementation of the force matching method.
It was originally designed to optimize interpolated pair and EAM \cite{Daw:1983:1285} potentials for model systems and metals, but has since been considerably expanded, which will be described below.
The code is parallelized over reference configurations (calculating the forces in each of them is completely independent of the others) using the Message Passing Interface.
While \potfit was designed to cooperate closely with the MD code IMD \cite{Stadler:1997:1131,Roth:2000:317} and shares a number of force calculation routines and file formats with that code, there are output routines for the popular LAMMPS package \cite{Plimpton:1995:1}.
Currently, \potfit can determine potentials for a wide range of solid state systems, both in tabulated and in analytic form, which is a unique trait to force matching codes currently available.
For further details about the original program, including theoretical background and implementation details, the reader is referred to \cite{Brommer:2008:}.

In any force matching code, evaluating a candidate potential by calculating the forces, stresses and energies it produces in the reference structures, is by far the computationally most demanding part of the code.
In comparison, all other computations are negligible.
This is accounted for in \potfit in two ways:
First, the individual force calculations are sped up as far as possible.
Second, the optimization methods are chosen to use as few force calculations as possible.
The latter approach is more difficult to realize in global optimization methods that are based on stochastically sampling the parameter space (such simulated annealing and differential evolution mentioned below).
As a consequence, a typical \potfit run can take anything between a few minutes and many hours.
It should be noted however, that this time is dwarfed by the other tasks associated with determining an effective potential: selection and calculation of reference data, as well as testing and validating the resulting potential.
Overall, creating a new effective potential with \potfit is a task on the order of weeks to months.

There are two independent parts in \potfit, a force routine and an optimization algorithm.
While the latter is responsible for the adjustment of the parameter set $\balpha$, the former calculates the quantities $S(\balpha)$ using a specific interaction model.
This separation of target function evaluation and parameter modification makes \potfit flexible; it is a comparatively simple process to add either an additional force calculation routine or a minimization method.
In the following, we will describe those two main parts of the \potfit program, with a particular emphasis on new developments.

\subsection{Optimization}
\label{sec:optimize}

The optimization is responsible for adjusting the parameter set $\balpha$ to minimize the target function $Z(\balpha)$ (\ref{eqn:sumofsquares}).
To this end, there are three distinct optimization algorithms implemented in \potfit.
Powell's algorithm \cite{Powell:1965:303} is a gradient-free conjugate-gradient-like optimization method that takes advantage of the particular form of the target function as a sum of squares.
This algorithm is efficient in the number of force calculations required to reach a local minimum in parameter space.
In contrast, a method based on simulated annealing \cite{Kirkpatrick:1983:671} is used to sample parts of parameter space in a Monte-Carlo based fashion.
The method outlined by Corana \emph{et al.}\ \cite{Corana:1987:262} is designed to operate on continuous variables and was successfully used particularly with tabulated potentials; a significant share of the work referenced in section \ref{sec:interact} below made use of this facility.
The implementation of these two methods in \potfit has been described in the original publication \cite{Brommer:2007:295}.
Differential evolution \cite{Storn:1997:341} is a variant of the genetic algorithm (GA) \cite{Goldberg:1989} approach, where a population of potentials is evolved to an optimal solution.
This method proved to be especially helpful for the optimization of analytic potentials (cf.\ section \ref{sec:potent-repr}) and is described in more detail below.
Simulated annealing and differential evolution are non-deterministic methods that unlike Powell's algorithm can leave the basin of attraction of a local minimum in favour of a potentially better solution to the optimization problem.

\subsubsection{Differential evolution}
\label{sec:de}
Genetic algorithms perform an optimisation by mimicking natural selection.
This is achieved by creating a population of candidate solutions to the optimisation problem (which in the case of force matching would be a group of trial parameter sets $\balpha$).
This population is then evolved from one generation to the next, using techniques inspired by biological evolution, like mutation or natural selection.
Typically ``weaker'' members of a population are replaced by ``fitter'' offspring created by combination of traits from other members of the population, thus increasing the overall quality of the population.
Here ``weak'' and``fit'' and refer to the optimisation criteria, which in force matching is the target function value.
The genetic algorithm implemented in \potfit is an example of differential evolution (DE), which is a class of a vector based stochastic optimization methods introduced by \cite{Storn:1997:341}.
There a population $P$ is used, with
\begin{eqnarray}
  P_g = \{\bm{x}_{i,g}\}, \quad i=0,1,\ldots,N_p-1, \quad g=0,1,\ldots,g_{max}
  \label{eqn:evo_pop1} \\
  \bm{x}_{i,g} = \{x_{j,i,g}\}, \quad j=0,1,\ldots,D-1, \label{eqn:evo_pop2}
\end{eqnarray}
where $N_p$ denotes the number of population vectors, $g$ counts the generations and $D$ is the dimensionality of the problem.
The vector $\bm{x}_{i,g}$ in \eref{eqn:evo_pop1} is equivalent to a set of parameters $\bm{\alpha}$ from \eref{eqn:sumofsquares}, with $D=\dim{\bm{\alpha}}$.

DE consists of three parts: mutation, crossover and selection.
For reasons of parallelization, DE works with two concurrent arrays for the population.
One holds the current and one the subsequent generation of parameters.
After an entire evolution step has been performed, the next generation is copied to the array holding the current one.
Afterwards the next step is started, unless some abortion criteria have been met.

\paragraph{Mutation} In the first part of the DE optimization loop a mutation vector $\bm{v}_{i,g}$ is created from the current population by combining several individuals $\bm{x}_{i,g}$.
There are different approaches on how to perform the combination, the basic version of DE uses the following:
\begin{equation}
  \bm{v}_{i,g} = \bm{x}_{r_0,g} + F\left(\bm{x}_{r_1,g}-\bm{x}_{r_2,g}\right).
  \label{eqn:DErand1bin}
\end{equation}
The indices $r_0$, $r_1$ and $r_2$ are mutually exclusive integer random numbers and $F$ a constant $\in\left(0,1\right]$.
Each mutation vector is composed of a random vector $\bm{x}_{r0,g}$ plus a weighted difference of two other random vectors $\bm{x}_{r1,g}$ and $\bm{x}_{r2,g}$.

\paragraph{Crossover} To get a diversity enhancement in the population, a crossover step is introduced.
It mixes the previously generated mutation vectors $\bm{v}_{i,g}$ with the population vectors $\bm{x}_{i,g}$ in order to generate trial vectors $\bm{u}_{i,g}$.
Generally just a binary choice is made, which is defined as
\begin{equation}
  \bm{u}_{i,g} = u_{j,i,g} =
  \cases{
    v_{j,i,g} \qquad \mathrm{if} \quad \mathrm{rand}_j[0,1)<C_r, \cr
    x_{j,i,g} \qquad \mathrm{otherwise}.
  }
\end{equation}
The $j$th component of the mutant vector is accepted for the trial vector with a probability of $C_r$.
If this is not the case, the component of the target vector is retained.
rand$_j[0,1)$ is a uniformly created random number and $C_r$ is called crossover rate.
In order to prevent the case $\bm{u}_{i,g}=\bm{x}_{i,g}$ at least one component of the mutant vector $\bm{v}_{i,g}$ is always accepted.
A general recommendation of $C_r\in[0.8,1.0]$ is given in \cite{Storn:1997:341}.

\paragraph{Selection} The last part of the DE optimization loop is a simple one-to-one selection.
Each trial vector $\bm{u}_{i,g}$ competes against the corresponding target vector $\bm{x}_{i,g}$.
The one which yields the lower target function $Z$ survives into the next generation $g+1$:
\begin{equation}
  \bm{x}_{i,g+1} =
  \cases{
    \bm{u}_{i,g} \quad \mathrm{if} \quad Z(\bm{u}_{i,g})\leq Z(\bm{x}_{i,g}), \cr
    \bm{x}_{i,g} \quad \mathrm{otherwise}.
  }
\end{equation}
After the selection has been performed the algorithm checks, if any of the abortion criteria are met.
If that is the case, the optimization is complete, otherwise another optimization loop will be performed.
A user defined critical threshold, with a default value of $10^{-6}$, is the main abortion criterion.
After each loop, the difference of the target vectors with the smallest and largest target function is calculated.
As long as this difference is greater than the critical threshold of $10^{-6}$ the optimization is continued.
That means, the algorithm will continue, until all target vectors have retracted to a very narrow volume in parameter space.

The DE variant implemented in \potfit is termed DE/rand/1/bin in \cite{Storn:1997:341}, as it mutates a random vector with one set of differences and accepts a crossover according to independent binomial experiments.
A pseudocode representation of this method can be found in the same reference \cite{Storn:1997:341}.

\subsection{Interaction models}
\label{sec:interact}
In its original implementation, \potfit could optimize pair and embedded atom method (EAM) \cite{Daw:1983:1285} potentials.
This functionality has been used extensively to create effective potentials for a wide range of mostly metallic materials:\footnote{This list is formed from potentials published in peer-reviewed articles in ISI indexed journals. For details about application range, reference data, etc., the reader is referred to the original publications.}
\begin{description}
\item[Metals and simple alloys] Nb \cite{Fellinger:2010:144119}, Al$_3$Ti\cite{Wang:2010:144203}, 14 face centered cubic (fcc) metals (Ag, Al, Au, Ca, Ce, Cu, Ir, Ni, Pb, Pd, Pt, Rh, Sr, Yb)\cite{Sheng:2011:134118}, U \cite{Smirnova:2012:107,Smirnova:2012:15702}, Au \cite{Starikov:2011:642}, Mg and Mg--Y\cite{Pei:2013:43020}, HfCo$_7$ \cite{Nguyen:2013:3281}. %
\item[Complex metallic alloys (CMAs)] Al--Ni--Co \cite{Brommer:2006:753}, Nb--Cr\cite{Roesch:2006:517} Ca--Cd\cite{Brommer:2007:2671} Mg--Zn\cite{Brommer:2009:97}, Al--Co\cite{Sonntag:2010:77}, Al--Pd--Mn\cite{Schopf:2012:54201}.
\item[Bulk metallic glasses (BMGs)] Cu--Zr(--Al) \cite{Cheng:2008:5263}, Pd--Si \cite{Ding:2012:60201}, Cu--Zr \cite{Zhang:2014:337}.
  \item[Metal--noble gas intercalation]  Mo--Xe \cite{Starikov:2011:104109}, Mo--U--Xe \cite{Smirnova:2013:35011}.
  \item[Transition metal chalcogenides (TMCs)] Cu$_2$Se, Cu$_2$Te \cite{Nguyen:2013:165502}.
  \item[Oxides] TiO$_2$ \cite{Wu:2014:35402}.
\end{description}

To extend the functionality of \potfit to different material classes, more potential models have been since been added to the program code.
This also motivated a further addition to the code that allows to directly optimize the parameters of a certain functional form (a so-called analytic potential), as not all of the new interaction models can be adequately represented in the tabulated form required by the original \potfit implementation.
Both these developments are described below.

\subsubsection{Potential representation}
\label{sec:potent-repr}

In an analytic potential, the functions creating the potential (e.g.\ a pair interaction $\phi(r_{ij})$) are defined by the parameters of a fixed functional form (e.g.\ $\phi(r_{ij})=A\exp(-\lambda r)$, with parameters $A$ and $\lambda$).
In contrast, tabulated potential functions are represented by their values at (not necessarily equidistant) sampling points.
The latter method avoids a model bias introduced by the selection of functional form, but typically requires many more parameters.
Also, in some cases, it may be possible to attach a physical interpretation to potential parameter (e.g.\ charge $q_i$ in Coulomb potential $q_iq_jr_{ij}^{-1}$).
Additionally, tabulated potentials require information in the reference data to support all sampling points.
This may be an issue with minority constituents of compound systems, where the respective pair distribution functions may have large gaps.
For analytic potentials, this is less an issue, as the values of the parameters affect a potential function globally.

To extend the functionality of \potfit, analytic potentials have been implemented.
Several predefined functions can be used to define a potential; additional functions can be implemented very easily.
A cutoff function $\Psi(r)$ is provided to make the potentials and their derivatives vanish at the cutoff distance $r_c$.
This functionality has first been used to determine EAM potentials for quasicrystal approximants in the Al--Pd--Mn \cite{Schopf:2012:54201} system.

\subsubsection{ADP potentials}
\label{sec:pot-adp}

Based on the EAM model, an angular dependent potential (ADP) \cite{Mishin:2005:4029} was introduced to account for directional forces.
Like EAM it is composed of purely pairwise contributions and does not contain three-body terms explicitly.
In an orthogonal Cartesian system the potential energy is given as
\begin{equation}
    E_{\mathrm{pot}} = \sum_{i<j}\phi_{ij}(r_{ij}) + \sum_iF_i(n_i)
  + \frac{1}{2}\sum_{i,\alpha}(\mu_i^\alpha)^2
  + \frac{1}{2}\sum_{i,\alpha,\beta}(\lambda_i^{\alpha\beta})^2
  - \frac{1}{6}\sum_i\nu_i^2.
  \label{eqn:adp}
\end{equation}
where indices $i$ and $j$ run over all atoms and superscripts $\alpha,\beta=1,2,3$ refer to Cartesian directions $x, y$ and $z$.
The first two terms in \eref{eqn:adp}, pair potential $\phi$ and embedding function $F$, are regular EAM terms.
An artificial electron density, $n_i=\sum_{j\neq i}\rho_j(r_{ij})$, is calculated as the sum of contributions from all neighbouring atoms using the transfer function $\rho_j$ of atom $j$.
The terms three to five in \eref{eqn:adp} are responsible for the indirect directional dependence of the potential.
The vectors $\mu_i$ and tensors $\lambda_i$ are functions of two additional pairwise potentials $u(r)$ and $w(r)$,
\begin{equation}
    \mu_i^\alpha = \sum_{j\neq i}u_{ij}(r_{ij})r_{ij}^\alpha, \qquad
    \lambda_i^{\alpha\beta}=\sum_{j\neq i}w_{ij}(r_{ij})r_{ij}^\alpha r_{ij}^\beta, \qquad
  \nu_i = \sum_\alpha\lambda_i^{\alpha\alpha},
\end{equation}
while $\nu_i$ is the trace of the $\lambda_i$-tensor.

Angular dependent potentials can be thought of as a kind of multipole expansion.
The terms $\mu_i$ and $\lambda_i$ are measures of the dipole and quadrupole distortion of the local environment of atom $i$.
For a perfect cubic symmetry they vanish, otherwise they increase the energy.

\subsubsection{Coulomb and dipole interactions}
Ionic solids can in general not be described adequately with short-ranged pair or EAM potentials.
The Coulomb interaction between charged particles falls of as $r^{-1}$ and there is no safe radius (i.e., converging to the true result) where the interaction can be cut off \cite{Wolf:1999:8254} -- which would be required in standard MD codes.
Thus, a force matching code must also treat Coulomb interactions differently from short-ranged pair potentials.
Additionally, it was shown that certain properties in various oxides cannot be reproduced correctly from pure pair potentials \cite{Beck:2012:485401}.
There, the polarizable oxygen model by Tangney and Scandolo (TS) \cite{Tangney:2002:8898} provides a much better description at limited computational cost.
In the TS model, the oxygen atoms are assigned a polarizability $\alpha$, leading to a dipole moment $\bm p$, which is determined self-consistently from the surrounding charges and dipoles, modified for short-range interactions.
The energy of the system is composed from the Coulomb interactions between charges and dipoles, and a short-ranged Morse-Stretch type interaction.

In densely charged systems such as ionic melts or solids, the long-range interactions between charges can be summed up using the linear scaling Wolf method \cite{Wolf:1999:8254}.
The same summation method can also be applied to the dipolar interactions in the TS method \cite{Brommer:2010:194109}, leading to a polarizable potential that scales linear in the number of particles.
This model has also been implemented in \potfit \cite{Beck:2011:234512}, and was used to generate effective interaction potentials for
SiO$_2$, MgO\cite{Beck:2011:234512}, Al$_2$O$_3$\cite{Hocker:2012:84707} and yttria-stabilized zirconia (YSZ) \cite{Iskandarov:2013:2811, Iskandarov:2015:15005}.
The Wolf summation is applied both in force matching and in MD simulation, which implies that the resulting potentials are consistent regarding the treatment of long-range interactions.

\subsubsection{Tersoff MOD potential and temperature dependent interactions}
\label{sec:temp-depend-inter}

One of the most commonly used empirical interatomic potentials in covalent materials is that developed by Tersoff \cite{Tersoff:1988:6991}.
In the original Tersoff interaction, the total potential energy $V$ is modelled as a sum of pairlike repulsive $V_R$ and attractive $V_A$ interactions with environment-dependent coefficient $b$:
\begin{eqnarray}
V & = & \frac{1}{2}\sum_{i\neq j}f_C(r_{ij})[V_R(r_{ij})-b_{ij}V_A(r_{ij})] \\
V_R & = & A\exp(-\lambda r_{ij}), \quad V_A = B\exp(-\mu r_{ij}) \\
b_{ij} &=& \left( 1 + \left(\zeta_{ij} \right)^{\eta} \right)^{-\delta} \\
\zeta_{ij} &=& \sum_{k(\neq i,j)} f_C(r_{ik})
g\left(\cos\theta\right) e^{\alpha(r_{ij}-r_{ik})^\beta}.
\end{eqnarray}
The modified angular-dependent term
\begin{equation}
g\left(\cos\theta\right) = c_1
+ \frac{c_2(h-\cos\theta)^2}{c_3+(h-\cos\theta)^2}\left[ 1 +
c_4 e^{-c_5(h-\cos\theta)^2}\right]
\end{equation}
and cutoff function
\begin{equation}
f_c (r) = \frac{1}{2} + \frac{9}{16} \cos \left( \pi \frac{r - R_1}{R_2 -
R_1} \right) - \frac{1}{16} \cos \left(3\pi \frac{r - R_1}{R_2 -
R_1} \right)
\end{equation}
were introduced in the modified Tersoff (MOD) \cite{Kumagai:2007:457} potential to improve the melting temperature value.

Under strong laser irradiation, the antibonding states of covalent materials become occupied at the expense of bonding states.
This corresponds to a non-thermal occupation of the electronic bands.
The electronic states thermalize over a few fs, albeit at an electronic temperature $T_e\gg T_l$ significantly above the lattice temperature $T_l$.%
\footnote{A non-thermal occupation of the electronic states is challenging to capture in standard DFT, which is why we assume a thermal occupation even for the first few timesteps.}
As a consequence, the potential energy surface and also the interatomic interactions change quasi instantaneously.
The resulting interatomic forces can induce nonthermal processes in the lattice such as melting or phase transformation.
This cannot be captured by standard potentials, as the forces acting on atoms in a structure representative of $T_l$ depend on the electronic temperature $T_e$.

To take these effects into account, some or all of the potential parameters can be made explicitly temperature dependent, e.g.\ $A=A(T_e)$.
Reference configurations are created  using finite-temperature density functional theory (FTDFT) \cite{Mermin:1965} calculations.
A set of parameters is then obtained for each temperature, where the temperature-independent parameters are kept fixed, and those dependent on the electronic temperature are allowed to vary smoothly in $T_e$ according to a sixth-order polynomial, i.e.,
\begin{equation}
  \label{eq:tdpar}
  A(T_e)=\sum_{k=0}^6 a_kT_e^k.
\end{equation}
The temperature-dependent modified Tersoff potential is called MOD*.
Starikov \emph{et al.}\ \cite{Starikov:2011:642,Norman:2012:792} used a similar approach to determine an electron-temperature dependent EAM potential for gold.
In a MD simulation, the time evolution and coupling between electron gas and lattice is then accounted for in a two-temperature model (TTM) \cite{Anisimov:1974:375}.
There the evolution of the electronic temperature is modelled in a continuum description, while the phononic system is represented by the point masses of the MD simulation.

\section{Validation and sample applications}
\label{sec:test}

The new features have been validated against calculations with the MD code IMD.
Here we present a review of recent results obtained with the newly implemented potential models.

\subsection{Angular dependent potentials for type I clathrates}
\label{sec:adp}

\begin{figure}[htbp]
  \includegraphics[width=0.2\textwidth]{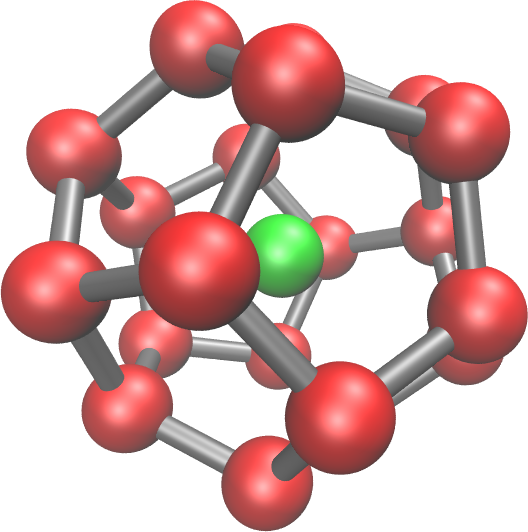}\hskip0.4cm
  \includegraphics[width=0.2\textwidth]{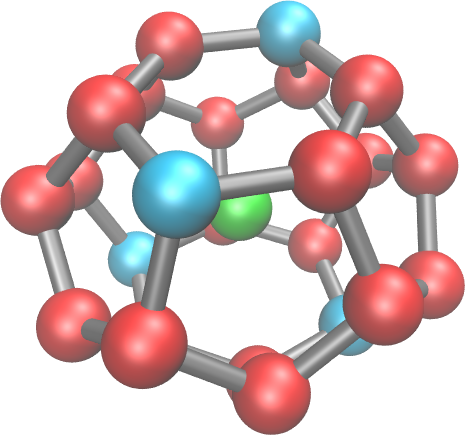}
  \caption{Schematic drawing of the different cage structures in the framework of type I
    clathrates. Left: Dodecahedron formed by 20 red atoms. The central atom is shown in green.
    Right: Tetrakaidecahedron formed by 24 red and teal atoms.
        From \cite{Schopf:2014:214306}.
    Copyright (2014) by the American Physical Society.}
  \label{fig:cages}
\end{figure}

To test the fitting of angular dependent potentials, semiconductor clathrate systems have been chosen.
Over the last few years they have become very promising candidates for thermoelectric applications, due to their vastly different transport coefficients for heat and electric current.
Type I clathrate structures are formed by two types of polyhedra, which consist of either Ge or Si atoms in our testcases.
Both are shown in Fig.~\ref{fig:cages}.
The smaller cages form a bcc lattice while the larger cages fill the remaining gaps.
More details on the clathrate structures are given in \cite{Karttunen:2011:1733}.

A second element, Ba, can be inserted into these cages, bringing the number of atoms per unit cell up to \num{54}.
These central atoms exhibit special phononic excitations, so called rattling modes.
Investigating the influence of these modes on macroscopic quantities requires simulations with thousands of atoms over several pico- or even nanoseconds.
Using \AI methods is not feasible for such simulations.
Molecular dynamics, however, using effective potentials, can give some insight into the influence of the rattling modes.

\begin{figure}[htpb]
  \includegraphics[width=.48\textwidth]{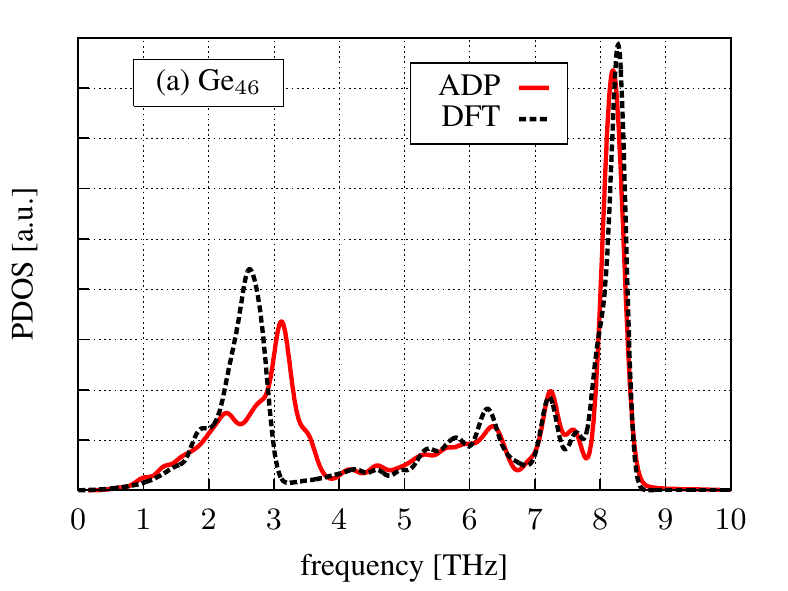}
  \includegraphics[width=.48\textwidth]{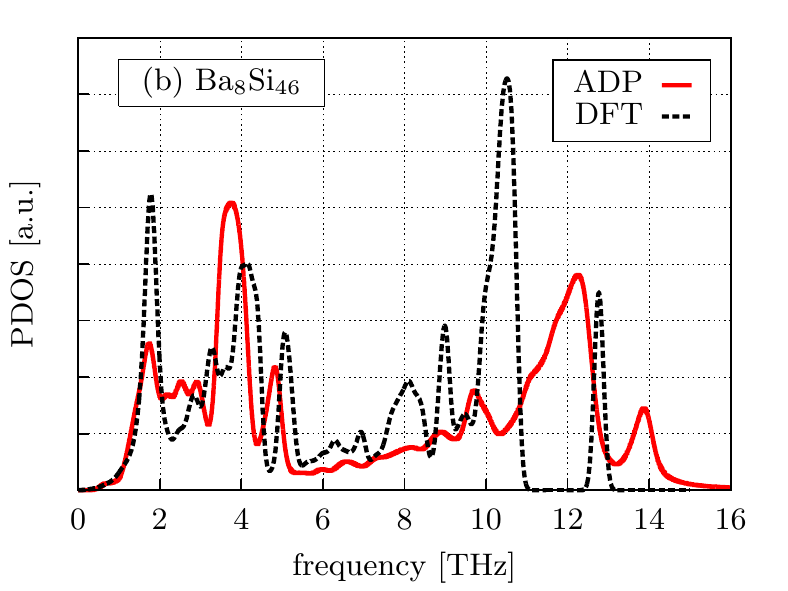}
  \caption{(a) Phonon density of states for the empty clathrate structure Ge$_{46}$. The ADP frequencies are scaled by a factor 1.02 to match the \AI data.
    (b) Phonon density of states for the Ba$_8$Si$_{46}$ clathrate structure. The ADP frequencies are unscaled.
    Images from \cite{Schopf:2014:214306}.
    Copyright (2014) by the American Physical Society.}
  \label{fig:pdos}
\end{figure}

ADP potentials have been fitted for the empty Ge cages as well as for Si cages filled with Ba atoms.
To this end, several static as well as dynamic \AI simulations have been performed to generate suitable reference configurations.
In total there were \num{6937} datapoints, i.e., forces, energies and stresses, for the Ge potential and \num{11393} for the BaSi potential.
Details of the selection and calculation of DFT values can be found in \cite{Schopf:2014:214306}.

A multi-stage approach was used for fitting the potential for the clathrate systems.
In a first step, the potentials for the framework atoms have been optimized and tested.
In a second run the additional potentials for the binary system were determined, while keeping the framework potential fixed.
To improve the results, a third run was performed, where all potentials were optimized simultaneously.
The chosen approach has the advantage, that the parameter space in the first two stages is significantly reduced when compared to a single-stage fitting process of all potentials.
Especially for systems with minor constituents or small amounts of impurities this multi-stage approach may be useful.

The clathrate structure can be well described with these potentials.
Static simulations showed a good agreement of the lattice constant with the \AI reference value as well as an experimental value.
The ADP potentials as well as \AI simulations yield a value of \SI{10.53}{\angstrom} for the Ge$_{46}$ system, the experimental value is \SI{10.55}{\angstrom}.
For the Ba$_8$Si$_{46}$ system the differences are a little larger, with \SI{10.18}{\angstrom} for the ADP potential, \SI{10.24}{\angstrom} for the \AI calculation and \SI{10.33}{\angstrom} for the experiment.

The designated application for the clathrate potentials are lattice dynamics.
They require a precise description of the phonons in the system.
To test the quality of the potentials the phonon density of states has been calculated with the effective potential and DFT methods, for comparison.

To calculate the phonon density of states, there are different approaches for \AI and MD calculations, which exploit the advantages of the respective calculation method.
The \AI results have been calculated with the \verb#phono.py#\cite{Togo:2008:134106} package, using the finite displacements method and the dynamical matrix approach.
For the MD simulations, the autocorrelation function of the particle trajectories has been used.

The results for the Ge$_{46}$ clathrate system are shown in Fig.~\ref{fig:pdos}(a).
The overall agreement of the ADP and \AI data is good.
For the ADP potentials the only major difference is the shift of the lowest peak.
The upper peak at \SI{8.2}{\thz} for Ge as well as the minor peaks in the intermediate frequency range are well reproduced.
For lower frequencies there are minor discrepancies, which are acceptable for effective potentials.
The oscillations of the ADP data for small frequencies can be attributed to anharmonicities in the potential, which are not present in the \AI calculations.

For the density of states of the Ba$_8$Si$_{46}$ system, shown in Fig.~\ref{fig:pdos}(b), the agreement for low frequencies is better than for high frequencies.
The three peaks at 2, 4, and \SI{5}{\thz} agree with only little deviations.
For frequencies above \SI{6}{\thz} the values of the ADP potential are constantly shifted about \SI{2}{\thz} to the right.
The reason for this is unclear. High frequencies correspond to the vibrations of the framework; they are also present in the empty cage structure Si$_{46}$.
Adding guest atoms creates phonon modes at low frequencies.

These calculations clearly show that the lattice dynamics of clathrate systems can be reproduced well by ADP potentials.
Further work has been performed  by determining the dynamical structure factor and the lattice thermal conductivity of these systems using the effective potentials.
All results have been published in \cite{Schopf:2014:214306}.

\subsection{Crack propagation in $\alpha$-alumina}
\label{sec:dipole}

Wolf-summated TS potentials were generated to study fracture in $\alpha$-alumina (Al$_2$O$_3$) \cite{Hocker:2012:84707}.
For a reasonable simulation of this complex process, the potential has to be able to adequately describe systems out of equilibrium, that are strained systems, and that contain free surfaces. 
Hence, the first challenge was to set up an \AI reference database which is able to represent all these system characteristics.
Finally, a reference database containing 67 structures with in total \num{24120} atoms was used, with atomic positions from three sources:
(i) strained structures (up to \SI{20}{\percent} strain along three different directions) at \SI{0}{\kelvin},
(ii) three distinct free surfaces in various terminations  \SI{0}{\kelvin}, and
(iii) snapshots from \AI MD trajectories at temperatures up to  \SI{2000}{\kelvin}.

The second challenge then was to validate wheter the potential is sufficently transferable and not overfitted (cf.\ Sec.~\ref{sec:inter-potent-from}).
Hence, before analyzing crack propahation, the resulting potential was validated to reproduce the basic properties of simple crystalline alumina (lattice constants, cohesive energies, vibrational properties), free surface characteristics (surface relaxations and surface energies) and the resulting stresses of strained configurations.
Potentials created with reference structures covering only one or two of the system classes mentioned above could not provide an adequate description of the system of interest: 
Those potentials failed the tests outlined here.
In this way, we arrived at the final reference data set iteratively; the validation procedure dictated expanding the reference data set to cover all three types of structures.

Mode I cracks were then simulated in  orthorhombic unit cells containing around \num{80000} atoms, using an elliptical seed crack.
A detailed description of the simulation conditions can be found in Ref. \cite{Hocker:2012:84707}.
The results are as expected strongly dependent on cleavage plane and to a lesser extent on propagation direction.
Cracks inserted in a $\{11\bar20\}$ plane propagate in both $[0\bar110]$ and $[0001]$ directions, with considerable disorder at the crack surface and atomic bridges across the crack.
In the dense   $\{0001\}$ planes, cracks do  not propagate in either $[\bar2110]$ or $[0\bar110]$ direction.
At higher energy release rates however, the $(0001)[0\bar110]$ crack diverts into a $\{10\bar12\}$ plane, which is one of the preferred cleavage planes.
Cracks inserted a $\{10\bar10\}$ plane moving in the $[0001]$ direction deviate slightly from the seed plane, propagating partially in a $\{10\bar12\}$ plane.
Both propagation directions in this plane leave behind atomic bridges across the crack surfaces.

In these simulations, additional insight can be gained from studying the induced dipole moment of the polarizable oxygen atoms
(Al atoms are assumed to possess no polarizability and consequently no dipole moment).
Two effects contribute to aligning the dipole moments: charged crack surfaces and material strain.
Charged surfaces occur depending on the cleavage plane, e.g.\ the oxygen-terminated $\{11\bar20\}$ surface show a dipole orientation normal to the crack surface.
More interestingly, polarization can also be induced by straining the material.
While bulk $\alpha$-Al$_2$O$_3$ is symmetric under inversion and thus shows no piezoelectricity (which simulations using the new potential reproduced), the crack breaks the inversion symmetry and shows macroscopic alignment of dipoles in inhomogeneously strained volumes of the simulation cell.

\begin{figure}
  \centering
  \includegraphics[width=.48\textwidth]{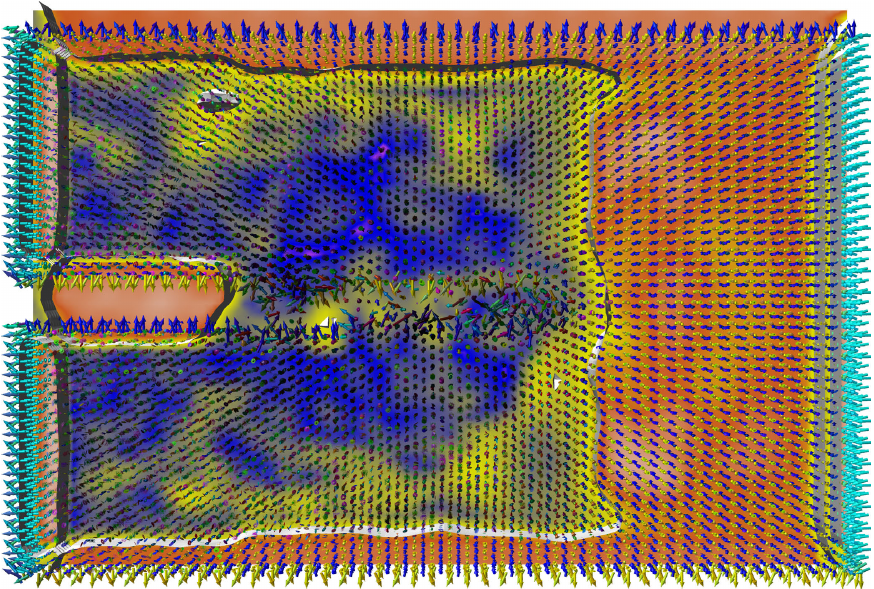}
\includegraphics[width=.48\textwidth]{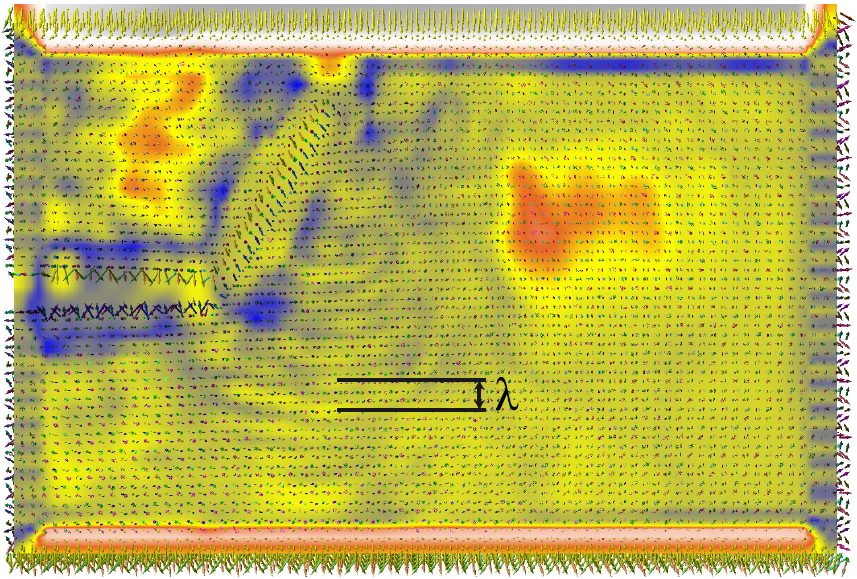}
  \caption{Left:
    Crack in the $(2\bar110)$ plane.
    Arrow glyphs represent O dipole moments, colour coded by direction. FA is encoded in the colourmap with red colours representing highly ordered dipole moments.
    The $\text{FA}=0.81$ isosurface separates the blue--yellow volume, where the passage of the crack has released the strain, from the highly ordered regions to the right that are still significantly strained.
Right: A crack in the $(0001)$ plane set up in the $[0\bar110]$ direction diverts into a $\{10\bar12\}$ plane.
The crack tip emits a circular wave of dipolar disorder with a wave length of about \SI{7.4}{\angstrom}.
Such a wave has not previously been observed. Reproduced with permission from \cite{Grottel:2012:2061}. \copyright\ 2012 IEEE.
}
  \label{fig:crack}
\end{figure}

To study the degree of dipolar ordering, a scalar quantity called fractional anisotropy (FA) \cite{Grottel:2012:2061} was calculated from the induced dipole moments in a simulation snapshot.
The FA represents how well aligned dipoles are over a neighbourhood corresponding to the potential cutoff radius (here: \SI{10}{\angstrom}), with 0 (1) corresponding to complete disorder (perfect order).
Two such visualizations can be found in figure \ref{fig:crack}, for two different cleavage planes.
There, the ordering of dipole moments serve as a probe of the local strain field, which serves to align neighbouring dipoles.

It should be noted, that the dipole moment calculated from a force-matched potential is primarily an empirical quantity, calculated from a polarizability chosen to optimally reproduce reference forces.
Dipole moments were not used as reference quantities.
However, the choice of potential model strongly implies that the vector orientations and magnitudes of this quantity can indeed be interpreted as dipole moments.
Additionally, the FA visualized in figure \ref{fig:crack} can be used as a generic order parameter reflecting on the local environment detached from a specific interpretation.

\subsection{Temperature dependent potential for Si}
\label{sec:temp-depend-potent}

We generated an electron-temperature dependent MOD* potential for silicon from FTDFT reference data calculated with VASP \cite{Kresse:1996:11169,Kresse:1993:558} using local density approximation and projector augmented waves \cite{Kresse:1999:1758}.
Using information from $44$ configurations -- simple cubic (sc), body-centered cubic (bcc), face-centered cubic (fcc) and cubic diamond crystal structures at $12$ different lattice constants each with in total $352$ atoms and electronic temperatures between \SI{0}{\kelvin} and \SI{25000}{\kelvin} -- we fit sixth order polynomials to the $A, B, \lambda_2$ and $\delta$ parameters, while keeping the remainder independent of temperature.
To test the resulting potential, we examined the electron temperature dependence of elastic constants and lattice constant for the represented crystal structures at low ionic temperatures.
This requires a non-equilibrium state of the electrons and ionic lattice, such as they occur after rapid material heating via strong laser fields.

According to the first-principles calculations used for MOD* fitting, the simple cubic crystal structure of silicon becomes energetically the most stable structure at high electronic temperatures above \SI{17500}{\kelvin}.
However, this transition cannot be observed directly in a MD simulation, because the electrons and lattice typically equilibrate after few picoseconds, which reduces the electronic temperature.
Alternatively, the stability of a structure can be assessed by its mechanical properties using the well-known Born stability criteria for the elastic constants $C_{ij}$.
In the particular case of a cubic crystal structure, the convexity of the free energy leads to the relations
\begin{equation}
  \label{eq:stability}
  C_{11} + 2C_{12} > 0, \; C_{44} > 0, \; C_{11} - C_{12} > 0.
\end{equation}

\begin{figure}
  \centering
  \includegraphics[width=.48\textwidth]{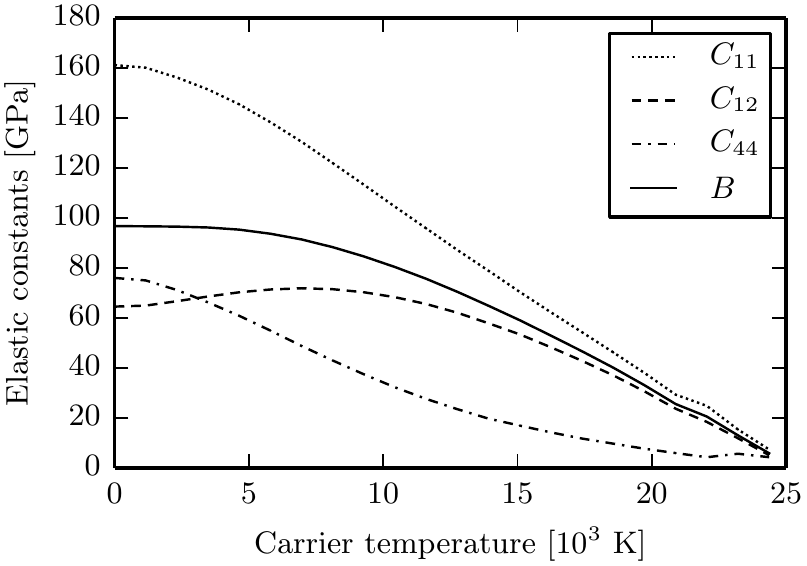}
\includegraphics[width=.48\textwidth]{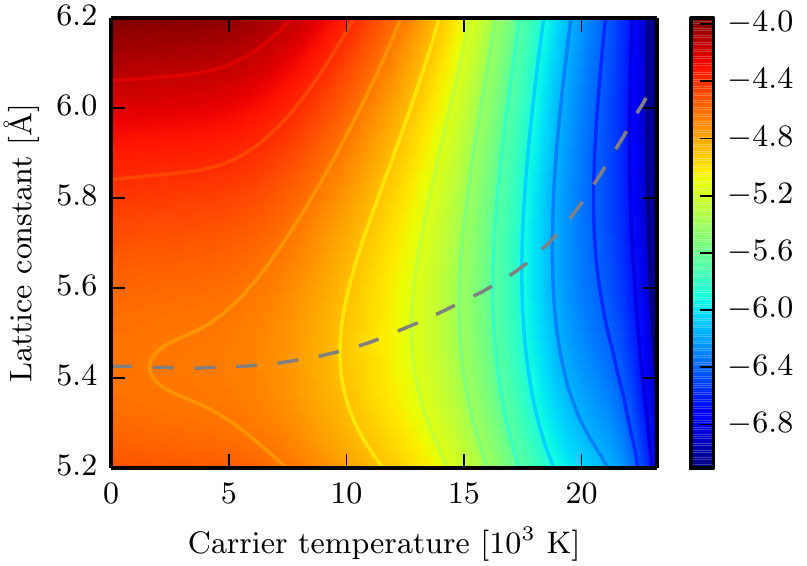}
  \caption{Electronic temperature dependence of elastic constants $C_{11},\,C_{12},\,C_{44}$, bulk modulus $B$ (left) and cohesive energies (right) of diamond silicon evaluated for different lattice constants using MOD* potential.}
  \label{fig:etd}
\end{figure}

Elastic constants $C_{11},\,C_{12},\,C_{44}$ and bulk moduli $B$ for diamond silicon obtained at different electronic temperatures using MOD* potential are shown in figure \ref{fig:etd} (left).
For comparison, the elastic constants $C_{11},\,C_{12},\,C_{44}$ for silicon at room temperature of \SI{167}{GPa}, \SI{65}{GPa} and \SI{81}{GPa} respectively are slightly underestimated by the MOD* potential.
The elastic constants and consequently the bulk modulus soften slowly initially and more rapidly with increasing carrier temperature, eventually vanishing and leading to structural transformation at electronic temperatures above \SI{25000}{\kelvin}.
The softening of interatomic forces is a direct consequence of electronic excitations to conduction bands, which break covalent bonds and induce more metallic behaviour in semiconductors and insulators after strong laser irradiation.

In the second test case we evaluated the cohesive energy of the bulk diamond silicon for a wide range of electronic temperatures and lattice constants. The results are shown in figure \ref{fig:etd} (right).
The room temperature lattice constant for silicon of \SI{5.43}{\angstrom}, which corresponds to an equilibrium bond length of \SI{2.35}{\angstrom}, and equilibrium cohesive energy of \SI{5.43}{eV} are well reproduced by MOD* potential.
With increasing electronic temperature we can observe initially slow and then rapid increase of the lattice constant, represented by the dashed line in the contour plot.
That expansion can be explained with a decrease of shared electronic number density and consequent increase of the repulsive forces between ions.

These initial tests show that the MOD* potential can capture essential effects of varying electronic temperature, which makes it suitable to simulate the effects of laser irradiation on silicon.
It is our aim to use the MOD* potential to study laser ablation of a Si film; this is ongoing work and will be presented at a later time.

\section{Conclusion}
\label{sec:conclude}
In this review article we demonstrate new features of the established force-matching code \potfit.
The results span a wide range of solid state materials, from ionic solids to metals and covalently bound materials of varying degree of complexity.
This demonstrates the flexibility and versatility of the program in determining interaction potentials for solid state systems.
By providing interfaces to standard DFT codes on the one hand, and several MD codes on the other, \potfit is an essential part of the sequential multiscale materials modelling stack.
In this way, it extends the length and time scales accessible to simulation for materials and conditions which up to now were limited to DFT methods.

One downside of effective interaction potentials is the lack of rigorous uncertainty quantification in the generation process: It is difficult to quantify the confidence of a certain property determined through MD simulation \emph{a priori}.
Only after the fact, an error can be assigned by comparison to experimental values.
For predictive modelling applications however, this is insufficient; the point is to yield reliable results including errorbars before an experiment is performed.
While there already exist approaches to use Bayesian techniques to estimate precision of potentials \cite{Frederiksen:2004:165501}, it would be a worthwhile goal to equip \potfit with the tools to provide not only an effective potential, but also a measure of the reliability of that result.

\section*{Acknowledgements}
The authors acknowledge funding from the Deutsche Forschungsgemeinschaft through Collaborative Research Project (SFB) 716 ``Dynamic simulation of systems with large particle numbers'', projects B.1 and B.5, and through Paketantrag PAK36 ``Physical Properties of Complex Metallic Alloys (PPCMA).'' Additional funding from EPSRC through grant no.\ P/L027682/1.
We would also like to thank all \potfit users for their feedback, comments, contributions and bug reports.

\section*{References}

\bibliographystyle{iop}
\bibliography{potfit,potfit_cites,forcematching}{}

\end{document}